\documentclass{article}
\begin{document}

\title{On the Leibniz bracket, the Schouten bracket\\ and the Laplacian}
\author{Bartolom\'{e} C{\sc oll}\\
Syst\`emes de r\'ef\'erence relativistes, SYRTE, Obsevatoire de Paris-CNRS,\\ 75014 Paris, France.\\ \texttt{bartolome.coll@obspm.fr}
\vspace{0.2cm}\\
Joan Josep F{\sc errando}\\
Departament d'Astronomia i Astrof\'{\i}sica, Universitat
de Val\`encia, \\46100 Burjassot, Val\`encia, Spain. \\ \texttt{joan.ferrando@uv.es}}
\maketitle
\baselineskip 18pt
\newtheorem{thm}{Theorem}
\newtheorem{cor}{Corollary}
\newtheorem{lem}{Lemma}
\newtheorem{prop}{Proposition}
\vspace{3cm}
\subsection*{Abstract}
The Leibniz bracket of an operator on a (graded) algebra is
defined and some of its properties are studied. A basic theorem
relating the Leibniz bracket of the commutator of two operators
to the Leibniz bracket of them, is obtained. Under some natural
conditions, the Leibniz bracket gives rise to a (graded) Lie
algebra structure. In particular, those algebras generated by
the Leibniz bracket of the divergence and the Laplacian
operators on the exterior algebra are considered, and the expression
of the Laplacian for the product of two functions is generalized for
arbitrary exterior forms.
\newpage
\section{Introduction}
\hspace{0.6cm}In mathematical physics, some operators of
interest are not derivations of the underlying algebraic
structures. Their complement to the Leibniz rule of derivation
defines then a product, called the Leibniz bracket. The Leibniz
bracket of a linear operator on an algebra is thus a bilinear
form that gives rise to a new algebra, called the Leibniz
algebra. Leibniz algebras present interesting properties, and
this work concerns them.

In particular, if the Leibniz bracket of an operator (its
adjoint action) is a derivation, the operator is of of degree
odd and its square vanishes or is also a derivation, then the
Leibniz bracket is a Lie bracket.

This is the case, for example, in the antibracket formalism context
\cite{wit}, for the exterior derivative considered as a second
order differential operator on the differential forms of finite
codimension: the antibracket can then be defined as the
corresponding Leibniz bracket, and some of its known properties
are simple consequences of the general results obtained here.

A similar situation occurs for the divergence operator over the
exterior algebra, for which the Leibniz bracket is nothing but
the Schouten \cite{sc1} \cite{sc2} bracket (in another different
context, an equivalent result has been obtained by Koszul
\cite{kos}). The expression obtained here relating the Schouten
bracket to the divergence operator is of interest in
mathematical physics. It allows, for example, to express Maxwell
equations in terms of Schouten bracket and to study {\it proper
variations} of Maxwell fields \cite{cf1} \cite{tolo}. It has
been also used to express the electromagnetic field equations in a
non linear theory which solves, in part, an old problem
concerning the existence and physical multiplicity of null
electromagnetic fields in general relativity \cite{tolo}
\cite{cf2}.
\par
The Leibniz bracket of the commutator of two operators admits a
simple expression: it is the commutator of the Leibniz bracket
of every one of them with respect to the operation defined by
the Leibniz bracket of the other one. For the Laplacian
operator, which {\it appears} as the (graded) commutator of the
divergence and the exterior derivative, the above expression may
be applied directly to it, giving the following interesting
result: the Leibniz bracket of the Laplace operator acting over
the exterior algebra equals the Leibniz bracket of the exterior
derivative acting over the Schouten algebra. This gives an
interesting generalization to the exterior algebra of the well
known expression for the Laplacian of a product of functions,
and it has been applied in the analysis of harmonic coordinates
in General Relativity \cite{tolo2}.

\section{Leibniz algebra of a graded operator}
\hspace{0.6cm} a) Let ${\mathcal E} = \bigoplus {\mathcal
E}_{a}$ be a commutative graded group and $\circ:{\mathcal E}
\times {\mathcal E} \rightarrow {\mathcal E}$ an operation
verifying ${\mathcal E}_{a} \circ {\mathcal E}_{b} \subseteq
{\mathcal E}_{a+b+k}$. Although it is always possible to
regraduate ${\mathcal E}$ so that k vanishes, we shall
retain the above graduation to avoid confusion when using
different operations, as we shall do; such a k will be called
the {\em degree} of the operation $ \circ $ (with respect to
this graduation).
\par
The known properties (and concepts) on a graded group ${\mathcal
E}$ concerning an operation $\circ$ of degree zero admit an equivalent form,
depending generically on the degree k, when an arbitrary
graduation is considered. Thus, the k-graded operation $ \circ $ is {\em commutative} (resp. {\em anticommutative}) if it verifies $A
\circ B = \epsilon (-1)^{(a+k)(b+k)} B \circ A$ with $\epsilon =
1$ (resp. $\epsilon = -1$), and it is {\em associative} if $A
\circ (B \circ C) = (A \circ B) \circ C$.

If ${\mathcal E}$ is a module
and $ \circ $ is bilinear, $({\mathcal E}, \circ )$ is said a
k-{\em graded algebra}. A {\em derivation of degree} r is a
r-graded endomorphism ${\bf D}$ on ${\mathcal E}$, ${\bf
D}({\mathcal E}_{a}) \subseteq {\mathcal E}_{a+r}$, verifying
the Leibniz rule ${\bf D}(A \circ B) = {\bf D}A \circ B +
(-1)^{(a+k)r}A \circ {\bf D}B$. An anticommutative k-graded
algebra $({\mathcal E},[,])$ verifying the Jacobi identity
\linebreak $\oint(-1)^{(a+k)(c+k)}[[A,B],C] = 0$ is said a
k-{\em graded Lie algebra}. Jacobi identity states,
equivalently, that the (a+k)-graded endomorphism ${\rm ad}A$,
${\rm ad}A(B)=[A,B]$, is a derivation on $({\mathcal E},[,])$.
If $({\mathcal E}, \circ )$ is a k-graded associative algebra,
the commutator defines a k-graded Lie algebra.
\par
Let ${\mathcal E}$ be a graded group, $ \circ $ a k-graded
operation and ${\bf P}$ a p-graded operator. When ${\bf P}$ does not satisfy the Leibniz rule, its "deviation" interests us. So we give the following definition: the {\em Leibniz
bracket} ${\mathcal L}_{{\bf P}}\langle \circ \rangle$ of ${\bf
P}$ with respect to $ \circ $ is the (p+k)-graded operation
given by:
\begin{equation}
{\mathcal L}_{{\bf P}}\langle \circ \rangle (A,B) = A \circ {\bf
P}(B) + (-1)^{p(a+k)}[{\bf P}(A) \circ B - {\bf P}(A \circ B)] \ .
\end{equation}
Of course, ${\bf P}$ verifies the Leibniz rule
iff the Leibniz
bracket ${\mathcal L}_{{\bf P}}\langle \circ \rangle$ vanishes
identically.

The Leibniz bracket of a linear operator with respect to a
bilinear operation is a bilinear operation, so that: {\em when
$({\mathcal E}, \circ )$ is a k-graded algebra and ${\bf P}$ is
a p-graded endomorphism, $({\mathcal E},{\mathcal L}_{{\bf
P}}\langle \circ \rangle )$ is a (k+p)-graded algebra}. We
call it the {\em Leibniz algebra} of ${\bf P}$ on $({\mathcal
E}, \circ )$. If ${\bf P}$ and ${\bf Q}$ are, respectively, p-
and q-graded endomorphisms, their commutator $[{\bf P}, {\bf Q}] =
{\bf P}{\bf Q} - (-1)^{pq}{\bf Q}{\bf P}$ is a (p+q)-graded
endomorphism. Then, taking into account that ${\mathcal L}_{{\bf
P}}\langle \circ \rangle $ and ${\mathcal L}_{{\bf Q}}\langle
\circ \rangle $ are respectively (p+k)- and (q+k)-graded
bilinear operations, and applying successively relation (1), one
obtains the fundamental result:
\newpage
\newtheorem{theo}{Theorem}
\begin{theo}
In a k-graded algebra $({\mathcal E}, \circ)$, the Leibniz
bracket of the commutator of two endomorphisms is related to the
Leibniz bracket of every one of them by
\begin{equation}
{\mathcal L}_{[{\bf P}, {\bf Q}]}\langle \circ \rangle  =
{\mathcal L}_{{\bf Q}}\langle {\mathcal L}_{{\bf P}}\langle
\circ \rangle \rangle - (-1)^{pq}{\mathcal L}_{{\bf P}}\langle
{\mathcal L}_{{\bf Q}}\langle \circ \rangle \rangle  \  ,
\end{equation}
\end{theo}
In Marx's style \cite{gm}: the Leibniz bracket
of the commutator $[{\bf P},{\bf Q}]$ of two endomorphisms ${\bf
P}$ and ${\bf Q}$ on the algebra $({\mathcal E}, \circ )$ equals
the graded difference between the Leibniz bracket of ${\bf Q}$
on the Leibniz algebra $({\mathcal E}, {\mathcal L}_{{\bf
P}}\langle \circ \rangle )$ of ${\bf P}$ and the Leibniz bracket of ${\bf P}$ on the Leibniz
algebra $({\mathcal E}, {\mathcal L}_{{\bf Q}}\langle \circ
\rangle )$ of ${\bf Q}$.

In particular, as ${\bf P}^{2}={\bf P}\cdot{\bf P}$ is a 2p-graded operator, it follows that
{\it for any odd-graded operator ${\bf P}$, one has}
\begin{equation}  \label{odd1}
{\mathcal L}_{{\bf P^{2}}}\langle \circ \rangle  = {\mathcal
L}_{{\bf P}}\langle {\mathcal L}_{{\bf P}}\langle \circ \rangle
\rangle   \  ,
\end{equation}
Let us note that ${\mathcal L}_{{\bf P}}\langle x \rangle $ may
be thought as {\em an operator ${\mathcal L}_{{\bf P}}$ over any
operation} $x$ {\em on} ${\mathcal E}$. In this sense, theorem 1 says that ${\mathcal L}_{[{\bf P},{\bf Q}]}\langle x \rangle  = [{\mathcal
L}_{{\bf Q}}, {\mathcal L}_{{\bf P}}]\langle x \rangle$, and relation (\ref{odd1}) says that
${\mathcal L}_{{\bf P^{2}}}\langle x \rangle  = ({\mathcal
L}_{{\bf P}})^{2}\langle x \rangle $.

Theorem 1 shows directly the well known result that if ${\bf P}$
and ${\bf Q}$ are derivations on $({\mathcal E}, \circ)$, so is
$[{\bf P},{\bf Q}]$. Also, from (\ref{odd1}), it follows:
\begin{lem}
The square ${\bf P}^{2}$ of an endomorphism ${\bf P}$ of odd
degree is a derivation on $({\mathcal E},  \circ )$ iff ${\bf
P}$ is a derivation on $({\mathcal E}, {\mathcal L}_{{\bf
P}}\langle \circ \rangle )$.
\end{lem}
On the other hand, if the operation $ \circ$ is commutative or anticommutative, i.e. $A \circ B = \epsilon (-1)^{(a+k)(b+k)} B \circ A$, one can find the following
result:
\begin{equation} \label{com}
{\mathcal L}_{{\bf P}}\langle \circ \rangle (A,B) = \epsilon
(-1)^{(a+k+p)(b+k+q)+p}{\mathcal L}_{{\bf P}}\langle \circ
\rangle(B,A)
\end{equation}
that is to say, {\it for a k-graded commutative (resp. anticommutative)
algebra $({\mathcal E}, \circ )$, the (k+p)-graded Leibniz
algebra $({\mathcal E},{\mathcal L}_{{\bf P}}\langle \circ
\rangle )$ is commutative (resp. anticommutative) if ${\bf P}$
is even-graded, and it is anticommutative (resp. commutative) if
${\bf P}$ is odd-graded}.

Let us denote, for simplicity, $\{A , B\}_{{\bf P}}={\mathcal
L}_{{\bf P}}\langle \circ \rangle (A,B)$. Then, when $({\mathcal E}, \circ)$ is a k-graded associative algebra
and ${\bf P}$ an endomorphism, one has
\begin{equation} \label{ad1}
\{A , B \circ C\}_{{\bf P}} - \{A , B\}_{{\bf P}} \circ C = (-1)^{p(b+k)}[\{A
\circ B, C\}_{{\bf P}} - A \circ \{B , C\}_{{\bf P}}]  \  .
\end{equation}
\vspace{0.1cm}
\par
b) Let $({\mathcal F}, \circ)$ be a 0-graded
associative and commutative algebra generated by its submodule
${\mathcal F}_{1}$, $\{,\}_{{\bf P}}$ be the Leibniz
bracket of the p-graded endomorphism ${\bf P}$ on $({\mathcal
F}, \circ)$, $\{A,B\}_{{\bf P}} \equiv {\mathcal L}_{{\bf
P}}\langle \circ \rangle (A,B)$, and  ${\rm ad}\{A\}_{{\bf P}}$
be the adjoint of A in the Leibniz algebra $({\mathcal
F},\{,\}_{{\bf P}})$, i.e. ${\rm ad}\{A\}_{{\bf P}}(B) \equiv
\{A,B\}_{{\bf P}}$.
\par
From the commutativity of $({\mathcal F}, \circ )$ and
relation (\ref{com}), equation (\ref{ad1}) may be written $\{C,B\}_{{\rm
ad}\{A\}_{{\bf P}}} = (-1)^{ca}\{A,B\}_{{\rm ad}\{C\}_{{\bf
P}}}$. Then, it follows: {\it for any p-graded endomorphism ${\bf
P}$ in $({\mathcal F}, \circ )$, one has}
\begin{equation} \label{ad2}
{\rm ad}\{C\}_{{\rm ad}\{A\}_{{\bf P}}} = (-1)^{ca+a+p}{\rm ad}\{A\}_{{\rm ad}\{C\}_{{\bf P}}}  \  .
\end{equation}
In particular, ${\rm ad}\{A\}_{{\bf P}}$ obeys the Leibniz rule
on the set $\{C\} \times {\mathcal F}$ iff ${\rm ad}\{C\}_{{\bf
P}}$ does it on the set $\{A\} \times {\mathcal F}$. Thus, iff
${\rm ad}\{X\}_{{\bf P}}$ is a derivation for every $X \in
{\mathcal F}_{1}$, ${\rm ad}\{A\}_{{\bf P}}$ verifies the
Leibniz rule on ${\mathcal F}_{1} \times {\mathcal F}$. But an
endomorphism that verifies the Leibniz rule on ${\mathcal F}_{1}
\circ {\mathcal F}$ is a derivation on $({\mathcal F}, \circ
)$, so that one has:
\begin{lem}
If ${\rm ad}\{X\}_{{\bf P}}$ is a derivation on $({\mathcal F},
\circ )$ for any $X$ of ${\mathcal F}_{1}$, then ${\rm
ad}\{A\}_{{\bf P}}$ is a derivation on $({\mathcal F}, \circ )$
for every $A$ of ${\mathcal F}$.
\end{lem}

If ${\bf P}$ is a derivation on its induced Leibniz algebra $({\mathcal
F}, \{,\}_{{\bf P}})$, the Leibniz rule may be written $[{\bf
P}, {\rm ad}\{A\}_{{\bf P}}] = {\rm ad}\{{\bf
P}(A)\}_{{\bf P}}$. Then, applying theorem 1 it follows that
${\rm ad}\{A\}_{{\bf P}}$ is a derivation on $({\mathcal
F},\{,\}_{{\bf P}})$ when ${\rm ad}\{A\}_{{\bf P}}$ and ${\rm
ad}\{{\bf P}(A)\}_{{\bf P}}$ are derivations on $({\mathcal F},
\circ )$. From this result and lemma 2 one has:
\begin{lem}
If ${\rm ad}\{X\}_{{\bf P}}$ is a derivation on $({\mathcal F},
\circ )$ for any $X \in {\mathcal F}_{1}$ and if ${\bf P}$ is a
derivation on $({\mathcal F},\{,\}_{{\bf P}})$, then ${\rm
ad}\{A\}_{{\bf P}}$ is a derivation on $({\mathcal F},\{,\}_{{\bf P}})$
for any $A \in {\mathcal F}$.
\end{lem}
For p odd, lemma 1 states that ${\bf P}$ is a derivation
on $({\mathcal F},\{,\}_{{\bf P}})$ iff ${\bf P}^{2}$ do it on
$({\mathcal F}, \circ )$. On the other hand, it follows from
relation (\ref{com}) that $({\mathcal F},\{,\}_{{\bf P}})$ is a
p-graded anticommutative algebra. But under this condition
Jacobi identity says equivalently that ${\rm ad}\{A\}_{{\bf P}}$
is a derivation on $({\mathcal F}, \{,\}_{{\bf P}})$. All that
and lemma 3 lead to the following result:
\begin{theo}
For p odd, if ${\bf P}^{2}$ and ${\rm ad}\{X\}_{{\bf P}}$, for
any $X$ in ${\mathcal F}_{1}$, are derivations on $({\mathcal
F}, \circ )$ then the Leibniz algebra $({\mathcal F},\{,\}_{{\bf
P}})$ is a p-graded Lie algebra.
\end{theo}
\section{Schouten bracket, divergence operator and Laplacian}
\hspace{0.6cm}a) Let $\Lambda^{p}$ (resp. $\Lambda^{\star p}$) be the set
of p-forms ({\em resp.} p-tensors) over the differential manifold
$M$, that is to say, the set of completely antisymmetric
covariant (resp. contravariant) tensor fields. Then,
$\Lambda  = \oplus \Lambda^{p}$ (resp. $\Lambda^{\star} =
\oplus \Lambda^{\star p}$) with the exterior product $\wedge$ is
a 0-graded associative and commutative algebra over the
function ring $\chi = \chi (M)$: the {\em exterior covariant algebra}
({\em resp.} {\em exterior contravariant algebra}). We shall denote by
$\alpha, \beta, \gamma$ the elements of $\Lambda $, and by $A, B, C$ those of $\Lambda^{\star}$,
with corresponding degrees $a, b, c$.
\par
Denote the interior product $i(A)\beta$, $[i(A)\beta]_{\underline{b-a}} = \frac{1}{a!}A^{\underline{a}}\beta_{\underline{a},\underline{b-a}}$ if $a \leq b$, by $(A, \beta)$ and put $(\beta,A) = (-1)^{a(b-a)}(A,\beta)$. When $X \in \Lambda^{1}$,
one has the usual interior product $i(X)$ which is a derivation of
degree -1 on $(\Lambda , \wedge)$. Moreover, one has
\begin{eqnarray}
(\gamma,A \wedge B) = ((A,\gamma),B) +
(-1)^{ab}((B,\gamma),A)  \label{sch1} \qquad \qquad & {\rm if} \ \ c = a+b-1, \\
(A \wedge B, \gamma) = (B, (A,\gamma)), \quad
(\gamma, A \wedge B) = ((\gamma,B),A) \qquad & {\rm if} \ \  c \geq a+b \ .   \label{sch2}
\end{eqnarray}

Suppose now that $M$ is a n-dimensional and oriented manifold,
and let $\eta$ be a (covariant) volume element, $\eta^{\star}$
being its (contravariant) dual: $\eta_{p,n-p}\eta^{\star p',n-p}
= \epsilon(n-p)^{-1}\delta_{p}^{p'}, \epsilon = \pm 1$. Then,
the Hodge operators are given by $\ast A = (\eta, A)$, $\ast
\alpha = (\eta^{\star},\alpha)$ and verify $\ast \ast A =
\epsilon(-1)^{a(n-a)}A$. Therefore, if $a+b \leq n$,
\begin{equation}   \label{sch3}
\ast(A \wedge B) =(\ast B,A), \qquad \qquad \ast A \wedge \beta = \ast (\beta,A) \ .
\end{equation}
The set of real numbers ${\mathcal R}$ being a sub-ring of the
set of functions $\chi$, $(\Lambda , \wedge)$ and
$(\Lambda^{\star}, \wedge)$ are $\chi$-algebras and ${\mathcal
R}$-algebras. The exterior differentiation $d$ is a 1-graded
${\mathcal R}$-derivation on $(\Lambda , \wedge)$, and the
codifferentiation (divergence up to sign) is a (-1)-graded
${\mathcal R}$-endomorphism given by $\delta = \epsilon
(-1)^{na}\ast d \ast$. Then, from (\ref{sch3}) it follows,
\begin{equation}   \label{sch4}
\delta (A,\beta) = (\delta A, \beta) + (-1)^{r}(A,d \beta), \qquad r = a-b >0  \ .
\end{equation}
\vspace{0.1cm}

b) It is known that for $X, Y \in
\Lambda^{\star 1}$, $\delta(X \wedge Y) = (\delta X) Y - (\delta
Y) X - {\mathcal L}_{X}Y$, so that the operator $\delta$ is not
a derivation on $(\Lambda^{\star}, \wedge)$. Thus, it is possible
to consider the Leibniz bracket $\{\ ,\ \}_{\delta}$ of the codifferential operator
on the exterior contravariant algebra $(\Lambda^{\star},
\wedge)$,
\begin{equation}   \label{sch5}
(-1)^{a}\{A, B\}_{\delta} = \delta A
\wedge B + (-1)^{a} A \wedge \delta B - \delta(A \wedge B) \  .
\end{equation}
Taking into account relations (\ref{sch1})(\ref{sch2}) and (\ref{sch4}),
it is not difficult to show that, for any (a+b-1)-form $\gamma$, one has
\begin{equation}  \label{sch6}
(-1)^{a}i(\{A, B\}_{\delta}) \gamma =
(d(\gamma,B),A) + (-1)^{ab}(d(\gamma,A),B) - (d \gamma , A
\wedge B)  \ .
\end{equation}
\hspace{0.6cm} The Schouten bracket $\{\ ,\ \}$ of two
contravariant tensors \cite{sc1} is a first order differential
concomitant that generalize the Lie derivative \cite{sc2}. For
p-tensors (antisymmetric contravariant tensors) this bracket is
defined by its action over the closed forms \cite{lich},
$i(\{A,B\})\gamma = (d(\gamma,B),A) + (-1)^{ab}(d(\gamma,A),B)$.
Comparing this relation and (\ref{sch6}), it follows $\{A , B\} =
(-1)^{a}\{A, B \}_{\delta} $, and one has the following
form of the Koszul \cite{kos} result:
\begin{theo}
The Schouten bracket is, up to a graded factor, the Leibniz bracket
of the operator $\delta$ on the exterior contravariant-algebra
$(\Lambda^{\star}, \wedge)$: $\{,\} = (-1)^{a}\{,\}_{\delta} $.
 Explicitly:
\begin{equation}   \label{sch7}
\{A,B\} =
\delta A \wedge B + (-1)^{a} A \wedge \delta B - \delta(A \wedge B),
\end{equation}
\end{theo}
This result justifies that we name {\em Leibniz-Schouten
bracket} the Leibniz bracket $\{,\}_{\delta} $ of the operator
$\delta$ on the exterior contravariant-algebra. Is is worth
pointing out that both, the Schouten bracket and the
Leibniz-Schouten bracket, define on the exterior
contravariant-algebra two {\em equivalent} structures of
(-1)-graded algebra, which we name, respectively, {\em Schouten
algebra} and {\em Leibniz-Schouten algebra}. Althought equivalent,
it is to be noted that the Schouten algebra does {\em not} satisfies
the standard writing of a Lie algebra properties, meanwhile the
Leibniz-Schouten algebra does. Let us see that.

It is not difficult to see that $\forall X
\in \Lambda^{\star 1}$, $\forall A \in \Lambda^{\star p}$, one
has $\{X,A\} = {\mathcal L}_{X}A$; that shows how the Schouten
bracket generalizes the Lie derivative. Let us write $\{A,B\} =
{\mathcal L}_{A}B$, $\forall A, B \in \Lambda^{\star}$; as it is
known, ${\mathcal L}_{X}, X \in \Lambda^{\star 1}$, is a
derivation and $\delta$ is a (-1)-graded endomorphism on the
0-graded associative and commutative algebra $(\Lambda^{\star
p}, \wedge)$ such that $\delta^{2}=0$. As a consequence, the
Leibniz-Schouten bracket $\{,\}_{\delta} $ satisfies the
hypothesis of theorem 2 and so {\em the Leibniz-Schouten algebra
$(\Lambda^{\star},\{,\}_{\delta} )$ is a (-1)-graded Lie
algebra}, that is, $\{\Lambda^{\star a}, \Lambda^{\star
b}\}_{\delta} \subseteq \Lambda^{\star a+b-1}$ and:
\begin{equation}  \label{lie}
\{A,B\}_{\delta} = -(-1)^{(a-1)(b-1)}\{B,A\}_{\delta}, \quad \quad
\oint(-1)^{(a-1)(c-1)}\{\{A,B\}_{\delta},C\}_{\delta} = 0  \  .
\end{equation}
The Schouten bracket $\{,\}$ also satisfies $\{\Lambda^{\star a},
\Lambda^{\star b}\} \subseteq \Lambda^{\star a+b-1}$, and the
properties of the Leibniz-Schouten Lie algebra (\ref{lie}) can equivalently be
written in terms of the Schouten bracket as:
\begin{equation}  \label{nolie}
\{A,B\} = (-1)^{ab}\{B,A\}, \qquad \qquad \oint(-1)^{ac}\{\{A,B\},C\} = 0  \  .
\end{equation}
Let us note that these last relations (\ref{nolie}) satisfied by the Schouten algebra
do not reduce, by any regraduation, to the standard ones of a Lie algebra.

Jacobi identity equivalently states, the following generalization for the Lie
derivative with respect to the Lie bracket:
\begin{displaymath}
{\mathcal L}_{\{A,B\}} = -(-1)^{a}[{\mathcal L}_{A},{\mathcal
L}_{B}]  \ .
\end{displaymath}
Also, from lemmas 1 and 2 and taking into account the
properties of the codifferential operator, it follows that: {\em
i) The codifferential operator $\delta$ is a ${\mathcal
R}$-derivation on the Leibniz-Schouten algebra:}
\begin{displaymath}
-\delta \{A,B\} = \{\delta A,B\} + (-1)^{a}\{A,\delta B\} \ ,
\end{displaymath}
{\em ii) The operator ${\mathcal L}_{A}$ is a ${\mathcal
R}$-derivation on the exterior contravariant algebra:}
\begin{displaymath}
{\mathcal L}_{A}(B \wedge C) = {\mathcal L}_{A}B \wedge C +
(-1)^{b(a-1)}B \wedge {\mathcal L}_{A}C  \ .
\end{displaymath}
The property i) gives the generalization of the
commutator of the codifferential and Lie derivative operators:
\begin{displaymath}
[\delta,{\mathcal L}_{A}] \equiv \delta {\mathcal L}_{A} +
(-1)^{a}{\mathcal L}_{A}\delta = -{\mathcal L}_{\delta A}  \ .
\end{displaymath}

On the other hand, equation (\ref{sch4}) may be written
$[i(\beta), \delta] = i(d\beta)$. But $i(\omega)$ is a
derivation on $(\Lambda^{\star}, \wedge)$ for any 1-form
$\omega$. Then, taking into account theorem 1, we have
${\mathcal L}_{i(\omega)}\langle \{\, \}_{\delta} \rangle =
{\mathcal L}_{i(d\omega)}\langle \wedge \rangle$. In particular,
when $\omega$ is a closed 1-form, then $i(\omega)$ is a
derivation on the Leibniz-Schouten algebra. \vspace{0.1cm}

c) Suppose now $M$ endowed with a (pseudo-)Riemannian metric
$g$, allowing to identify $(\Lambda , \wedge)$ and
$(\Lambda^{\star}, \wedge)$. The Laplacian operator is then the
{\em graded} commutator of the differential and codifferential
operators:
\begin{displaymath}
\Delta = [d,\delta] \equiv d\delta + \delta d  \  .
\end{displaymath}
It is known that $\Delta$ is not a derivation on the exterior algebra. From theorem 1
its Leibniz bracket is given by:
\begin{theo}
The Leibniz bracket of the Laplacian operator on the exterior
algebra equals the Leibniz bracket of the exterior derivative on
the Leibniz-Schouten algebra: ${\mathcal L}_{\Delta}\langle \wedge
\rangle = {\mathcal L}_{d}\langle \{\, \}_{\delta} \rangle$. Explicitly:
\begin{equation}   \label{sch8}
\Delta \alpha \wedge \beta + \alpha \wedge \Delta \beta -
\Delta(\alpha \wedge \beta) = \{d\alpha, \beta\} + (-1)^{a} \{\alpha,d\beta\} + d\{\alpha,\beta\}
\end{equation}
where $\alpha$ and $\beta$ are arbitrary $a$- and $b$-forms respectively.
\end{theo}
This theorem gives the generalization to the exterior algebra of
the expression for the Laplacian of a product of functions:
$\Delta f . h + f . \Delta h - \Delta (f.h) = 2 (df, dh)$.

\end{document}